# $Mg_yNi_{1-y}(H_x)$ thin films deposited by magnetron co-sputtering

T. Mongstad[a,*], C. C. You[a], A. Thogersen[a], J. P. Maehlen[a], Ch. Platzer-Björkman[b], B. C. Hauback[a], S. Zh. Karazhanov[a]

[a] Institute for Energy Technology, P.O. Box 40, NO-2027 Kjeller, Norway
[b] Uppsala University, Solid State Electronics, Box 534, SE-751 21 Uppsala

Abstract:
In this work we have synthesised thin films of $Mg_yNi_{1-y}(H_x)$ metal and metal hydride with $y$ between 0 and 1. The films are deposited by magnetron co-sputtering of metallic targets of Mg and Ni. Metallic $Mg_yNi_{1-y}$ films were deposited with pure Ar plasma while $Mg_yNi_{1-y}H_x$ hydride films were deposited reactively with 30% $H_2$ in the Ar plasma. The depositions were done with a fixed substrate carrier, producing films with a spatial gradient in the Mg and Ni composition. The combinatorial method of co-sputtering gives an insight into the phase diagram of $Mg_yNi_{1-y}$ and $Mg_yNi_{1-y}H_x$, and allows us to investigate structural, optical and electrical properties of the resulting alloys. Our results show that reactive sputtering gives direct deposition of metal hydride films, with high purity in the case of $Mg_{\sim2}NiH_{\sim4}$. We have observed limited oxidation after several months of exposure to ambient conditions. $Mg_yNi_{1-y}$ and $Mg_yNi_{1-y}H_x$ films might be applied for optical control in smart windows, optical sensors and as a semiconducting material for photovoltaic solar cells.

Keywords: metal hydrides, semiconductors, thin films, vapour deposition, optical spectroscopy, transmission electron microscopy (TEM)

Graphical abstract:

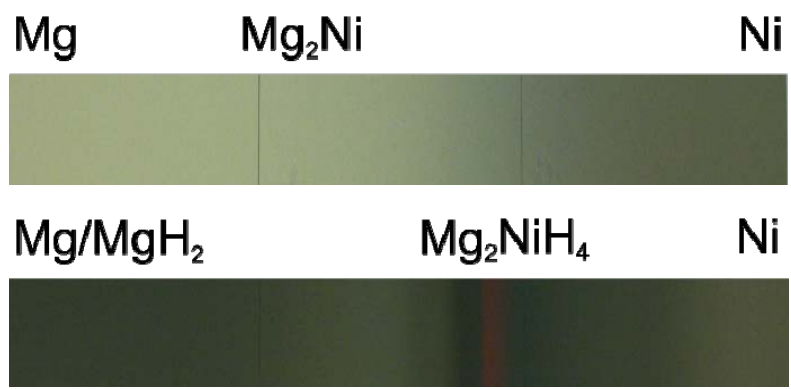

* Corresponding author. Tel.: +47 99228200; Fax: +47 63899964; E-mail address: trygve.mongstad@ife.no.

# 1. Introduction

The Mg-Ni-H system has been heavily investigated for many years, mainly because of the large capacity for hydrogen storage. $MgH_2$ and $Mg_2NiH_4$ stores respectively 7.6 wt% and 3.6 wt% hydrogen [1]. Also several other applications have been proposed, involving thin films of Mg-Ni-H for use in smart windows [2], hydrogen sensors [3], semiconductor electronics [4], solar cells [5] and data storage [6]. These applications rely on the respectively insulating and semiconducting nature of the hydrides of Mg and $Mg_2Ni$. The band gaps of $MgH_2$ and $Mg_2NiH_4$ are reported to be respectively 5.6 eV [7] and 1.7 eV [8].

There are several paths to synthesising Mg-Ni hydrides. Hydrogenation of metallic powders is the conventional method for synthesis of bulk $MgH_2$ and $Mg_2NiH_4$. For $Mg_yNi_{1-y}H_x$ thin films, metallic Mg-Ni films are normally deposited by magnetron sputtering or evaporation and hydrogenated *ex-situ* after the deposition. A few nanometers of Pd capping is commonly used in order to protect the films from oxidation and as a catalyst to accelerate the hydrogen uptake [9], but also films without Pd cap layers can be hydrogenated under exposure to hydrogen for several hours [6,10]. Direct formation of metal hydride films *in-situ* can be performed by activated reactive evaporation [11,12] or reactive sputtering deposition as applied in the current work.

Reactive sputtering is a versatile technique, a relatively simple and very efficient process suitable for up-scaling to industrial processing. Direct deposition of hydride films does not require the deposition of any catalytic metal layers to form the hydride. This is preferable when investigating optical and electrical properties due to the conductivity and opacity of catalytic metallic films as for example Pd. Direct deposition can also be a way to avoid stress resulting from the volume expansion of hydrogenation of metal films [13]. Reactive sputter deposition is an uncommon way of synthesising metal hydrides, but has earlier been applied for films of $TiH_2$ [14], $CaH_2$ [15], NaH [15], $MgH_2$ [16], $NaAlH_4$ [17] and $YH_x$ [18]. Direct deposition of Mg-Ni metal hydrides by reactive co-sputtering has not been reported earlier.

# 2. Experimental details

## 2.1 Thin film deposition

The thin films were deposited by sputtering metallic Mg and Ni targets in a Leybold Optics A550V7 inline sputtering system. The purity of the targets was 99.5% and 99.8%, respectively. The target area was 125 x 600 $mm^2$ and the targets were set at an angle of 30º with respect to the substrate against each other in order to enhance the co-deposition. The mean distance from the targets to the substrate was 116 mm, and the centre-to-centre distance between the targets was 210 mm. The Mg target was operated with RF (13.56 MHz) power and Ni with DC power, varying the power on each target from 150 to 1000 W. 1000 W corresponds to a power density of 1.3 $W/cm^2$ at the target. For the samples described in the results section, the power applied for the Mg/Ni targets was 600W/600W for the metallic films and 800W/200W for the reactively deposited hydride films. The purity of the gases was 5N for Ar and 6N for $H_2$. The base pressure of the chamber was $1.6 \times 10^{-4}$ Pa, and the depositions were performed under 0.4 Pa pressure. A 3:7 mixing ratio of $H_2$ to Ar was used for the reactive depositions with hydrogen. The total gas flow was 200 sccm. The depositions were carried out at room temperature, but the substrate was slightly heated due to the deposition power. Thus the temperature increased with the deposition time, and could reach ~50-100

°C after one hour of deposition at 1000W. The samples were deposited on glass substrates (Menzel-Gläser microscope slides) for energy-dispersive X-ray spectroscopy (EDS) and measurement of electrical, optical and structural parameters, on polished Si substrates for transmission electron microscopy (TEM), on glassy carbon substrates for Rutherford backscattering (RBS) experiments, and on aluminium foil for thermal desorption spectroscopy (TDS). The substrates were cleaned in de-ionized water in an ultrasound bath for 15 minutes and blow-dried with pressurised $N_2$. The 76 × 26 mm$^2$ glass substrates were placed in a row with 5 substrates after each other to collect samples from a 380 mm long region with a continuous gradient in the composition going from almost pure Mg ($y \approx 1$) to almost pure Ni ($y \approx 0$). The deposition rates were estimated by measuring the thickness and weight of thin film samples ex-situ. The thickness of the deposited films was determined by stylus profilometry. The mass of the substrate was measured before deposition and the compared to the weight of the substrate with the deposited film after deposition in order to measure the mass of the film. The typical mass of a film was 2-10 mg with an accuracy of 0.1 mg.

## 2.2 Sample characterization

Electrical resistivity was measured by using a 4-point probe in collinear geometry. The resistivity was measured every 5 mm on 380 mm long compositional gradient samples, and with the measurement lines not electrically insulated from each other. The optical reflection and transmission were measured using an Ocean Optics QE65000 spectrometer for ultraviolet and visible light and an Ocean Optics NIRQUEST for the near infrared range. The optical measurements were done with the light incident from the film to air interface. Optical measurements on compositional gradient samples were performed by measuring the reflection and transmission on samples every 10 mm. The diameter of the probe light beam was 5 mm. Structural characterization was performed by grazing incidence X-ray diffraction (GI-XRD, 0.5° incident angle) in a Bruker D8 Discover with Cu-Kα radiation ($\lambda$ =1.5409 Å). Samples with a spatial gradient in the composition $y$ were characterized by GI-XRD on points 15 mm apart, corresponding to the width of the x-ray beam. 15 mm on the gradient sample corresponds to a $\Delta y \approx 0.1$ close to $Mg_{0.5}Ni_{0.5}(H_x)$, and down to $\Delta y \approx 0.01$ for the positions with close to pure Mg or Ni. Cross-sectional TEM samples were prepared by ion milling using a Gatan precision ion polishing system with 5 kV gun voltage. The samples were analysed by high resolution TEM (HRTEM) in a 200 keV JEOL 2010 F microscope with a Gatan imaging filter and detector. The spherical ($C_s$) and chromatic aberration ($C_c$) coefficients of the objective lens were 0.5 and 1.1 mm, respectively. The point to point resolution was 0.194 nm at Scherzer focus ( - 42 nm). RBS measurements were performed at the Tandem Accelerator Laboratory at Uppsala University, Sweden. The RBS data was analysed using the SIMNRA software [19]. EDS spectra were collected in a Hitachi S-4800 scanning electron microscope with Noran System SIX and analysed using the software NSS 3.0. TDS was performed to evaluate the hydrogen desorption properties. Thin film samples deposited on aluminium foil were heated in dynamic vacuum with a constant heating rate of 5 K/min. The hydrogen evolution was measured using a calibrated vacuum sensor consisting of a cold cathode ionization gauge (inverted magnetron) and a Pirani gauge. To remove moisture, the films were degassed for several hours at ca 40 °C prior to the measurement.

# 3. The sputtering deposition process

The deposition rates of Mg and Ni were estimated in order to investigate how the deposition rates are affected under co-sputtering and sputtering with $H_2$ as a reactive gas. The thickness deposition rate $r_t$ (nm/min) and the gravimetric deposition rate $r_g$ (g/cm$^2$/min) was estimated by post-deposition measurements of thickness and mass, respectively. The mass is directly related to the composition and number of atoms in the sample, and is thus a better measurement when using sputter rates to estimate the chemical composition of a co-sputtered film.

In the case of co-sputtering of metallic films, the calibration of single-target sputter rates are often used to determine the composition of deposited films [6,20]. The deposition rate is measured for single target sputtering, and it is assumed that the deposition rate of each element is not changed under co-sputtering. Fig. 1(a) demonstrates that this assumption is satisfying for non-reactive co-sputtering of Mg and Ni. This observation was confirmed by measurements of the chemical composition by EDS. The deposition rate is enhanced by 5-15% for both targets under co-sputtering, but the sputter rates of each element are not severely affected by co-sputtering.

Fig. 1(b) compares the gravimetric deposition rates for single-target and for co-sputtering in the case of reactive deposition with 30% $H_2$ in the Ar plasma during deposition. The depositions were done with 800 W RF power on the Mg target and 200 W DC power on the Ni target. By comparing Fig. 1(a) and Fig. 1(b), we can make the following three observations; Firstly, the Ni deposition rate is not notably affected by adding $H_2$ to the plasma. Secondly, we see that the Mg single-target deposition rate is reduced slightly (~15%) when introducing $H_2$. Thirdly, we see that while co-sputtering in the non-reactive case does not give substantial change in the position-dependent individual elemental deposition rates, co-sputtering with reactive $H_2$ in the plasma gives a very notable decrease in the Mg yield. The Ni deposition rate is not substantially affected even under reactive co-sputtering. Under reactive sputtering deposition the single-target deposition rates are not conserved for co-sputtering, and the individual rates can therefore not be used to estimate the composition of a reactive co-sputter deposited film.

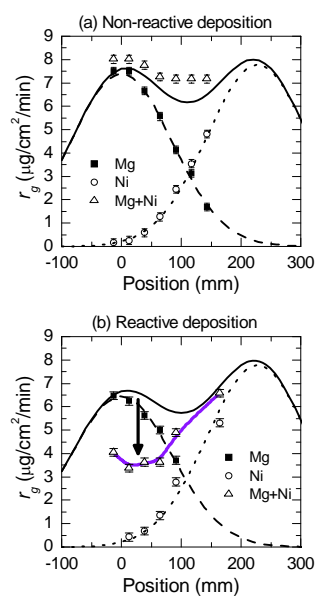

*Fig. 1. Measurement and fit of gravimetric deposition rates for single target deposition of Mg and Ni, and for co-sputtering of Mg and Ni. (a) Non-reactive deposition and (b) reactive deposition with 30% $H_2$ in the plasma. The full-drawn black line demonstrates the sum of the Gaussian fits to the separate single target deposition rates. The purple line is displayed as a guide to the eye for the measurement of the reactive co-sputtering deposition rates in (b), which together with the arrow demonstrate the quenching of the Mg deposition rate under reactive co-sputtering.*

Since individual sputter rates does not give a good estimation of the chemical composition of reactively co-sputtered $Mg_yNi_{1-y}H_x$ films, EDS and RBS was used to find the composition. Results from EDS and RBS gave results that agreed within an error of a few percent in the composition number $y$. The hydrogen content of the samples was not estimated.

## 4. Material properties of the deposited films

### 4.1 Electrical resistivity of deposited films

Fig. 2 displays the electrical resistivity as a function of the composition $y$ in $Mg_yNi_{1-y}(H_x)$ samples deposited with and without reactive hydrogen. The resistivity obtained for reactively deposited hydride films was several orders of magnitude higher than that obtained for metallic films, suggesting formation of semiconducting $Mg_2NiH_4$ and $MgH_2$.

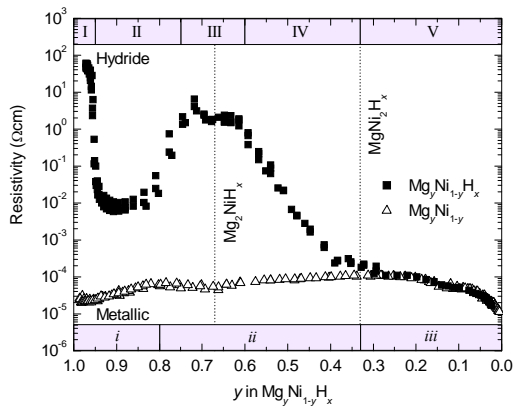

Fig. 2. Resistivity as a function of composition for $Mg_yNi_{1-y}(H_x)$ samples deposited without and with reactive $H_2$.

As a guide in the further discussion, we can define three regions in the compositional gradient of the $Mg_yNi_{1-y}$ film deposited without $H_2$:
  i. $1.0 > y > 0.80$: Mg with solid solution of Ni
  ii. $0.80 > y > 0.33$: $Mg_2Ni$ with solid solution of Mg or Ni
  iii. $0.33 > y > 0.0$: Ni with solid solution of Mg

The two borders between the three regions are marked by maxima in the resistivity. Similar maxima were also observed for $Mg_yNi_{1-y}$ by Gremaud et al., and were explained by disorder scattering [21]. The minima for the metallic films, on the extreme right and left of Fig. 2, correspond to Mg and Ni metal with small amounts of Ni and Mg, respectively. The resistivity for the Mg-rich side approaches 25 μΩcm, and 15 μΩcm on the Ni-rich side. Literature values for resistivity of bulk metal at room temperature are 4.5 μΩcm for Mg and 7.0 μΩcm for Ni [22].

For the reactive deposition with $H_2$ the signs of hydride formation in the resistivity are clearly shown in Fig. 2. We can define five regions in the compositional gradient of the $Mg_yNi_{1-y}H_x$ film;
  I. $1.0 > y > 0.95$: $MgH_2$ with Mg particle inclusions and solid solution of Ni
  II. $0.95 > y > 0.75$: Mg with $MgH_2$ particle inclusions and solid solution of Ni
  III. $0.75 > y > 0.60$: $Mg_2NiH_4$ with solid solution of Mg or Ni
  IV. $0.60 > y > 0.33$: $Mg_2NiH_4$ with Ni or $MgNi_2$ particle inclusions
  V. $0.33 > y > 0.0$: Ni with solid solution of Mg

Region I has high resistivity in the order of 10-100 Ωcm, corresponding to the formation of $MgH_2$, probably with inclusions of metallic Mg particles. Pure $MgH_2$ is an insulator, and we have in earlier work found that reactive sputter deposition of only magnesium produced $MgH_x$ films with resistivity at least 2 orders of magnitude higher than we found in the current work [16]. In region II, with increasing amount of Ni, there is a region with a resistivity of around 10 mΩcm. The XRD data presented later proves formation of a substantial amount of crystalline Mg in this region, and thus explains the lower resistivity than in the surrounding regions I and III. It seems that the added Ni limits the formation of $MgH_2$ in region II. The resistivity in this area is still 2-3 orders of magnitude higher than what was found for similar compositions in the metallic $Mg_yNi_{1-y}$, suggesting a certain level of hydrogen incorporation modulating the resistivity. Furthermore, the low optical reflectivity (Fig. 3(b)) suggests presence of a substantial amount of hydrogen. The composition in region III corresponds to $y = 0.60 - 0.75$, in other words close to the $Mg_2NiH_4$ composition. Region III is characterized by a resistivity of 1-10 Ωcm. The high resistivity in this region suggests formation of semiconducting magnesium nickel hydride. Region IV shows a gradual decrease in resistivity with more and more metallic behavior as the amount of Ni is increased. In region V the composition changes from $MgNi_2(H_x)$ to Ni, and shows low resistivity similar to that of $Mg_yNi_{1-y}$ gradients deposited without $H_2$. From the resistivity it therefore appears that very little H is dissolved in this region.

By depositing films with substrate temperatures of up to ~100 ºC, the resistivity in the $Mg_2NiH_4$-region increased by 1-2 orders of magnitude. Longer deposition times of up to one hour also heat the substrate, and gave a similar increase in resistivity. The highest resistivity obtained for high temperature deposited samples was of ~400 Ωcm in the semiconducting region (region III). Enache et al. found a maximum resistivity of only 12.9 mΩcm for $Mg_2NiH_4$ films prepared by hydrogenation of Pd-capped metallic films [23]. Westerwaal et al, studying *in-situ* deposited films of $Mg_2NiH_4$ by activated reactive evaporation, found a considerably higher resistivity of 0.34 Ωcm [12], comparable to what we in the current work found for films deposited at room temperature (1-10 Ωcm). The resistivity maximum found by Westerwaal et al. was limited by the maximum power of their atomic hydrogen source, and we believe it could in principle have been increased further. Westerwaal et al. explained the deviation in the results found for *in-situ* deposited films and *ex-situ* hydrogenated films by the higher grain boundary density they found in the *in-situ* prepared films. In our opinion one could also question if it is possible in the experiments of Enache et al. to decouple the effect of the Pd cap layer in the resistivity measurements on Pd-capped thin $Mg_2NiH_4$ film. Both the conductivity of the Pd film and Pd islands as well as in-diffusion of Pd could give reduced resistance. An alternative explanation for the deviance in the resistivity between *ex-situ* hydrogenated and the *in-situ* deposited films is that the *in-situ* deposited films are closer to stoichiometry in H, approaching closer to the $Mg_2NiH_4$ composition. Hydrogen vacancies in $Mg_2NiH_4$ behave as n-type doping [23] that provides increased conductivity in the case of H under-stoichiometry. A hydride closer to H-stoichiometric $Mg_2NiH_4$ will have higher resistivity.

*4.2 Optical properties*
Fig. 3 shows the optical properties of the metallic and hydride films as a function of the composition $y$ in $Mg_yNi_{1-y}(H_x)$ and the wavelength of the light. The optical transmission of the metallic films deposited without $H_2$ is not displayed since it was close to zero for the thickness and wavelength ranges considered in this work. Fig. 3(a) displays the optical reflection spectra for the metallic films over the wavelength range 300-1700 nm. The reflection of the metallic samples is highest in the regions with the most pure metal. Pure Mg is a metal with very high reflectivity of close to 95% over the displayed wavelength range. Even though the reflectivity of the most Mg-rich region *i* is the highest, the solid solution of Ni lowers the reflection substantially, especially for shorter wavelengths. The reflection is further reduced in region *ii* as the Ni content in Mg is increasing, reaching a minimum for the MgNi$_2$ composition, which also corresponds to the maximum resistivity. There is a second minimum in region *iii* for Ni with 5% Mg. The reflectivity for Ni with little Mg content approaches very close to the reflectivity for pure Ni as calculated from known optical constants [24]. The reflection and transmission spectra for films deposited with reactive H$_2$ are displayed in Fig. 3(b) and (c), respectively. The thickness of the samples was 350 – 800 nm. The reflection spectra demonstrate the interesting optical properties of the Mg-Ni-H system. Region I shows very low reflection of 10-25 % for all measured wavelengths. Region II shows slightly higher reflection, but still much lower than what would be expected for a metal. The low reflection of unsaturated metal hydride films is a well-known phenomenon and has been referred to as the black state of metal hydride films. It has been explained by a co-existence of conducting metal grains and semiconducting or insulating metal hydride grains [25]. Region I containing mainly MgH$_2$ shows some transparency. The large optical band gap of pure MgH$_2$ makes it transparent for visible and infrared (IR) radiation. Region II shows no transmission, suggesting formation of mainly metallic Mg.

The most pronounced feature in the optical spectra of the hydride film (Fig. 3(b) and 3(c)) is the transparency in region III, close to the Mg$_2$NiH$_x$ composition. This can be explained by the formation of semiconducting Mg$_2$NiH$_4$. The reflection and transmission of Mg$_{2.1}$NiH$_4$ is displayed in Fig. 4(a). The high transparency suggests that no or little metallic Mg or Ni particles exist in this region, contrary to what is found when MgH$_2$ is formed in a reactive sputtering process [16]. The transparency reaches up to 75% as an average from the band gap ($\lambda = 800$ nm) and up to 1600 nm. The absorption coefficient can be estimated from the reflection and transmission data using the approximation $\alpha = 1/d \ln[(1-R)/T]$, where $d$ is the thickness of the film, and $R$ and $T$ are the measured reflection and transmission, respectively. The band gap can be estimated from the relation $\alpha h v = (h v - E_g)^m$, where $m = 2$ for amorphous semiconductors [26]. The band gap is then indicated by identifying the linear region of the plot of $(\alpha h\nu)^{1/2}$ as a function of the photon energy $h\nu$, as displayed in Fig. 4(b) using the data in Fig. 4(a). Here we find a band gap of 1.55 eV for $y = 0.68$ ($Mg_{2.1}NiH_x$). The band gap increases with increasing the Mg content and decreases with decreasing Mg content.

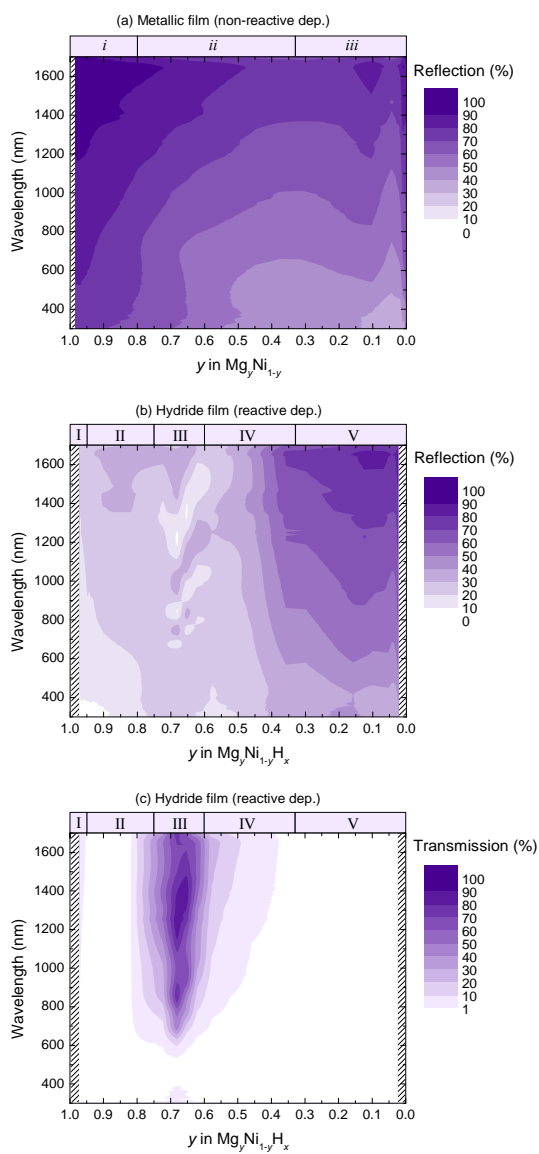

*Fig. 3. Optical properties of Mg$_y$Ni$_{1-y}$(H$_x$) films. (a) Reflection spectra for metallic Mg$_y$Ni$_{1-y}$ film deposited without reactive H$_2$. (b) Reflection and (c) transmission spectra for Mg$_y$Ni$_{1-y}$H$_x$ film deposited with reactive H$_2$. The thickness of the films was from 800 nm in the most Mg-rich region to 350 nm in the most Ni-rich region.*

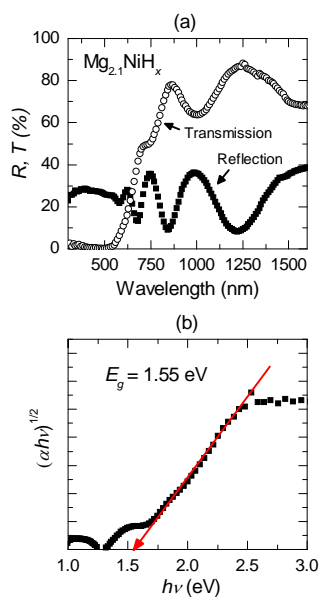

*Fig. 4. Optical properties and band gap assignment of a 505 nm thick amorphous $Mg_{2.1}NiH_x$ film ($y \approx 0.68$). (a) Optical reflection and transmission, and (b) band gap assignment plot for the same data showing band gap of 1.55 eV.*

In region IV we see a gradual reduction of the transparency and increase of the reflectivity as more metallic Ni becomes present. In region V, the most Ni-rich side, the optical spectra of the reactively deposited $Mg_yNi_{1-y}H_x$ film shows formation of pure Ni with high reflectivity and zero transmission. The reflectivity of region V in Fig. 3(b) shows the same features as that of region *iii* for the metallic film in Fig. 3(a).

### 4.3 Structural properties

The compositional gradient films were investigated by XRD to identify crystalline phases of metals and metal hydrides. The diffraction spectra for metallic $Mg_yNi_{1-y}$ are given in Fig. 5(a). The Bragg peaks corresponding to crystalline Mg, $Mg_2Ni$ and Ni are easily identifiable within the respective regions *i*, *ii* and *iii*. Mg crystallizes in a hexagonal space group ($P 6_3/m m c$, $a = 3.21$ Å, $c = 5.21$ Å) with the (002) reflection visible at 34.4° and the (103) reflection at 62.9°; $Mg_2Ni$ in a hexagonal space group ($P 6_2 2 2$, $a = 5.19$ Å, $c = 13.21$ Å) with the reflections (003) at 20.13° and (006) at 40.8°; Ni in a cubic space group ($F m - 3 m$, $a = 3.52$ Å) with the reflections (111) at 44.6°, (200) at 52.1° and (220) at 76.6°. In the region where crystalline Ni forms, the peak position for the (111) reflection moves from 44.6° to 44.2° as the Mg content increases. This corresponds to an expansion of the unit cell with changes from $a = 3.52$ Å in pure Ni to $a = 3.55$ Å for solid-solution Ni-Mg with 15-20% Mg. In between each crystalline region the film is amorphous. The amorphous regions correspond to the compositions with maxima in resistivity, see Fig. 2, and thus support the disorder scattering proposed by Gremaud et al [21].

XRD data for $Mg_yNi_{1-y}H_x$ films deposited with reactive $H_2$ is displayed in Fig. 5(b). Bragg peaks from $MgH_2$, Mg and Ni can be identified. $MgH_2$ is found in the tetragonal structure ($P 4_2/m n m$, $a = 4.52$, $c = 3.02$) with Bragg peaks from (110), (200) and (211) at 28.1°, 40.0° and 54.5°, respectively. In region I crystalline $MgH_2$ forms, including a small amount of crystalline Mg. This is similar to what we have earlier observed for pure Mg deposited reactively with $H_2$ [16]. In region II, diffraction from both crystalline Mg and $MgH_2$ is visible. In region III, the diffraction pattern shows no clear peaks, suggesting an amorphous structure. There are two halos centered at 23° and 40°, similarly to what has been observed for *ex-situ* hydrogenated $Mg_2Ni$ films [10] and nano-grained $Mg_2NiH_4$ [27]. In region IV, as the Ni content in $Mg_2NiH_4$ increases, the halo at 40° moves towards 44° and in region V the halo at ~44° disappears when crystalline Ni forms. The crystal structure in region V resembles the structure in region *iii* for samples deposited without reactive hydrogen, forming amorphous $MgNi_2$ and cubic Ni with Mg in solid solution.

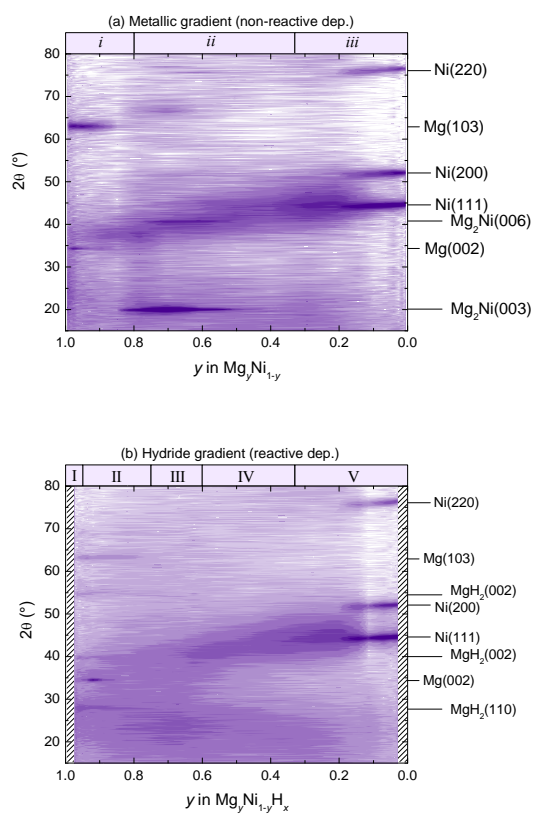

*Fig. 5. XRD spectra plotted as logarithmic intensity plots (counts per second), as a function of composition for (a) a metallic $Mg_yNi_{1-y}$ film deposited without reactive $H_2$ and (b) for a $Mg_yNi_{1-y}H_x$ film deposited with reactive $H_2$.*

To closer investigate the structure in the XRD-amorphous semiconducting region of the reactively deposited film (region III, $Mg_{\sim2}NiH_x$) a cross-sectional sample was prepared for TEM. Fig. 6(a) shows a TEM image of the 1 μm thick film with electron diffraction patterns from eight points through the thickness of the film.

The TEM image and diffraction patterns show a crystalline film. The diffraction pattern presented in ¤A is from the Si substrate. The diffraction patterns of the film found in area ¤B through ¤H showed diffraction from pure Mg, $Mg_2Ni$, Ni and MgO, as well as a monoclinic phase we believe belong to $Mg_2NiH_4$. Figure 9 shows the diffraction pattern of area ¤E in Figure 8, where the rings added to the image indicate lattice parameters belonging to monoclinic $Mg_2NiH_4$ and hexagonal $Mg_2Ni$. Of the possible $Mg_yNi_{1-y}$ and $Mg_yNi_{1-y}H_x$ phases, the reflections have the closest fit to the monoclinic $Mg_2NiH_4$ (*C 1 2 / c 1*) with *a* = 14.363 Å, *b* = 6.4052 Å, and *c* = 6.4963 Å (α= γ = 90º and β = 113.622º), known as the low-temperature (LT) structure of $Mg_2NiH_4$ [28]. Fig. 6(b) shows a HRTEM image of area ¤E in Fig. 6(a), demonstrating a grain size of 5-10 nm. No clusters of either pure Mg and/or Ni could be observed using HRTEM or EDS mapping of the film. Halos from amorphous MgO are observable mainly near the substrate interface and the surface of the film. The samples were not protected from air under the transfer from the ion milling chamber to the TEM apparatus, and can have been partly oxidized during the transfer. The film reacted slightly with the electron beam, which could lead to oxidation due to residual oxygen (O) in the TEM. Slight oxidation of the sample or crystallization of the film could also have occurred during sample preparation. It remains an open question whether the sample was crystalline originally, or if amorphous $Mg_2NiH_4$ crystallized under the effect of the energetic electron beam or the TEM sample preparation. The presence of metallic particles could also result from electron beam exposure, as hydrogen could have been released from the sample under interaction with the beam in vacuum..

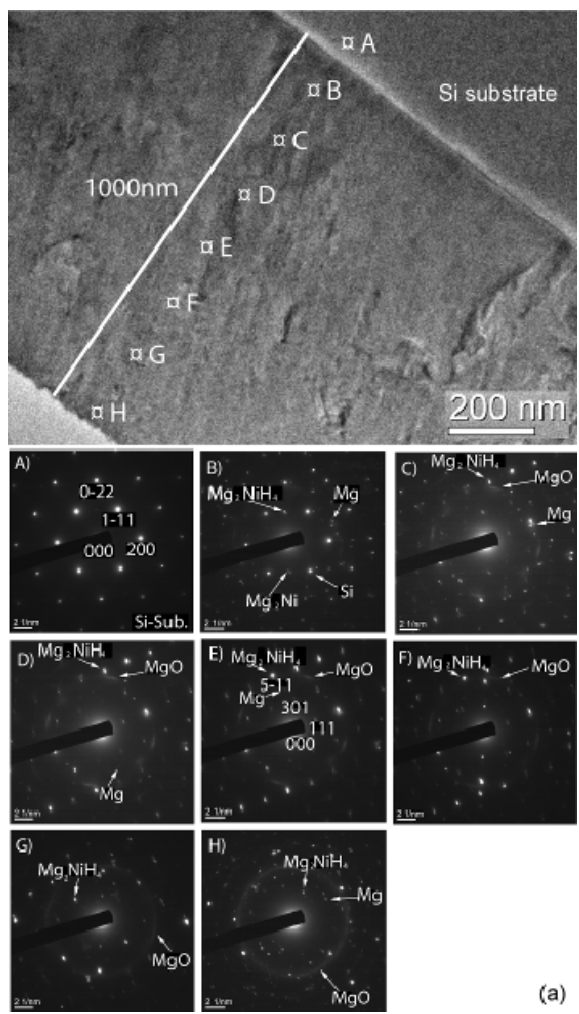

*Fig. 6. TEM image and electron diffraction patterns of a semiconducting $Mg_{\sim 2}NiH_x$ film (region III). (a) TEM image of a cross-section of a 1 μm thick film with electron diffraction patterns from different depths of the sample. The planes corresponding to some of the reflections of Si and $Mg_2NiH_4$ are indicated on A) and E), respectively. (b) HR-TEM showing the crystalline grains of $Mg_2NiH_4$.*

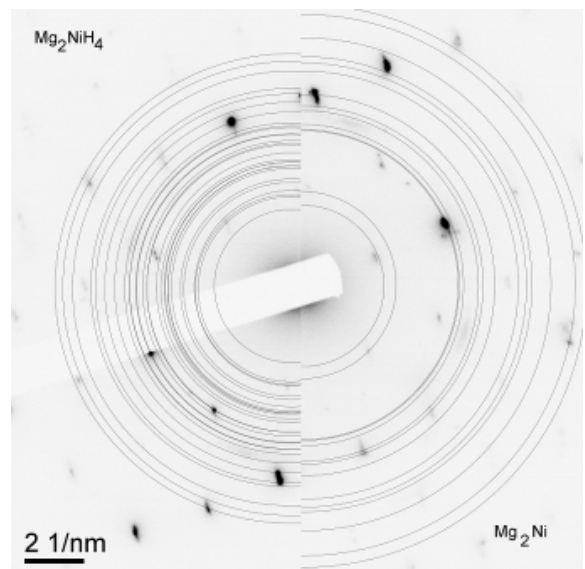

*Figure 7: The electron diffraction pattern of area ¤E in Figure 6. The rings indicate lattice parameters from the monoclinic $Mg_2NiH_4$ and the $Mg_2Ni$ phase.*

### 4.4 Chemical stability

Oxidation and dehydrogenation of metal hydrides can take place at ambient conditions. In our earlier work we observed that $MgH_2$ films deposited at room temperature by reactive sputtering were oxidized when stored at ambient conditions. In the current work, $Mg_yNi_{1-y}H_x$ films deposited with reactive $H_2$ were found to be reasonably stable in air. The resistivity and optical properties showed in Fig. 2 and Fig. 3 were checked several months after deposition, and no significant changes could be observed.

Fig. 8 shows the RBS spectra collected for different compositions for reactively deposited $Mg_yNi_{1-y}H_x$. The samples were transferred to the RBS chamber without any surface protection against oxidation in air. The table in Fig. 8 displays the composition as at% of each element calculated for each spectrum, and shows that the O content in these samples is relatively low. Hydrogen is ignored in the calculation because RBS is incapable of determining H concentrations. The sensitivity for O in RBS is low due to the low atomic mass of O. The O signal is low as compared to the noise level, and the O content thus has relative uncertainty of up to 50 %. The O is mainly located on the substrate interface and the surface of the film, as illustrated by the inset on Fig. 8. The stability of reactively deposited $Mg_yNi_{1-y}H_x$ against oxidation in air has also been observed earlier [12], and was explained by the formation of a surface oxide that protects the sample from further oxidation and dehydrogenation.

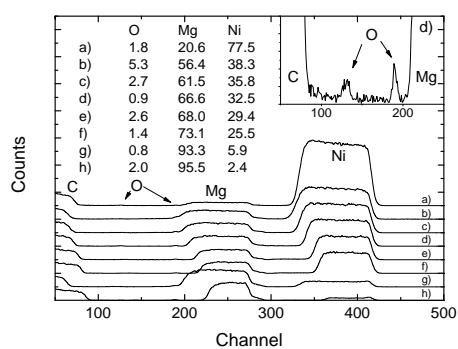

*Fig. 8. RBS spectra for different positions in a $Mg_yNi_{1-y}H_x$ film deposited with reactive $H_2$. The spectra demonstrate variations in Mg and Ni content, as well as a limited incorporation of O. The two O peaks on the inset demonstrates the two layers of high O content, one on the upper surface and the other at the substrate interface.*

Fig. 9 shows the TDS spectra for two samples deposited with reactive $H_2$ with different Mg-Ni composition. The sample with composition corresponding to $Mg_{\sim 2}NiH_x$ ($y \approx 0.67$) shows a major desorption peak at 240 ºC, the desorption starting at 200 ºC. There is also a small desorption peak at 375 ºC. The more Mg-rich sample with $y \approx 0.9$ shows no desorption until the large and narrow peak at 380 ºC, starting at 350 ºC. We attribute the peak at 240 ºC to desorption of H from $Mg_2NiH_4$ and the peak at 380 ºC to $MgH_2$. The positions of the desorption peaks are in agreement with results by Zaluski et al. on nano-grained $MgH_2$, $Mg_2NiH_4$ and mixtures of these [27].

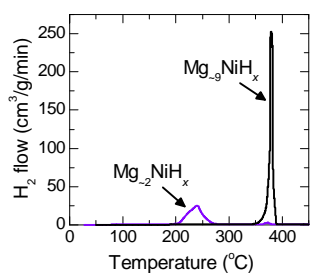

*Fig. 9. TDS spectra from thin film samples with $y \approx 0.67$ and $y \approx 0.9$.*

### 4.5 Thin films of $Mg_{\sim 2}NiH_{\sim 4}$ for solar cell applications

In the introduction, we shortly described the various applications that are suggested for Mg and Ni-based hydrides. In addition to the much studied switchable windows based on these materials [2], we mentioned the recent suggestion of utilising thin film $Mg_2NiH_4$ as a light-absorbing material for solar cells [4]. A suitable band gap, good electrical transport properties and long-term stability are the main requirements for solar cell semiconductors. The ideal band gap for a solar cell is around 1.4 eV [29], which is close to our estimation of $E_g = 1.55$ eV for the amorphous $Mg_{2.1}NiH_x$ (Fig. 4(b)). The electrical transport properties involving parameters as minority carrier life time and carrier mobility have not yet been studied. Furthermore, long term stability has not been tested under realistic conditions, but the fact that unprotected thin film samples are resistant against oxidation and that dehydrogenation only occurs at temperatures higher than 200 ºC (in vacuum) is promising with respect to this requirement. To summarize, the current results show that $Mg_{\sim 2}NiH_{\sim 4}$ is a promising candidate for a future solar cell material based on the abundant elements Mg, Ni and H, and that reactive sputter deposition is a suitable technique for deposition of thin films of semiconducting $Mg_{\sim 2}NiH_x$.

## 5. Conclusions

We have here given a description of thin film Mg and Ni co-sputter deposition without and with reactive $H_2$. The electrical, optical and structural properties of $Mg_yNi_{1-y}(H_x)$ films with compositional gradient have been discussed. The main findings are:

1. Single-target deposition rates can be used in order to estimate the composition of co-sputter deposited films of Mg and Ni without reactive $H_2$, but cannot be used for reactive co-sputter deposition of Mg and Ni with $H_2$.
2. Co-sputtered Mg and Ni react with $H_2$ in the reactive sputtering deposition process to form films of metal and metal hydride depending on the mixing ratio of Mg and Ni. Ni-rich compositions with y < 0.33 do not form hydride in this process.
3. Semiconducting $Mg_{\sim2}NiH_{\sim4}$ forms for 0.6 < y < 0.75 when sputtering with reactive $H_2$. Optical and electrical measurements show low presence of metallic particles in the as-deposited semiconducting hydride films. Monoclinic $Mg_2NiH_4$ was observed by TEM, with a grain size of 5-10 nm.
4. Mg with high Ni content does under reactive deposition with $H_2$ form a mixture of $MgH_2$ and Mg with relatively low resistivity and low optical reflection.
5. High stability against oxidation is observed for all compositions of reactively deposited $Mg_yNi_{1-y}H_x$.
6. $Mg_{\sim2}NiH_x$ films deposited by reactive sputtering with $H_2$ desorb hydrogen at 200-240 ºC.

The method of reactive sputter deposition is well-suited for studies of thin film metal hydrides based on Mg and Ni, especially for deposition of thin films of semiconducting $Mg_{\sim2}NiH_{\sim4}$.

## Acknowledgements

This work has been funded by the Research Council of Norway through the project "Thin and highly efficient silicon-based solar cells incorporating nanostructures," NFR Project No. 181884/S10. We thank Göran Possnert (Uppsala University) for help with the RBS measurements.